# Weak antilocalization effect of high-mobility two-dimensional electron gas in inversion layer on p-type HgCdTe

Rui Yang,[1] Kuanghong Gao,[1,2] Laiming Wei,[1] Xinzhi Liu,[1] Guolin Yu,[1,*] Tie Lin,[1]

Shaoling Guo,[1] Ning Dai,[1] and Junhao Chu[1,2]

[1]*National Laboratory for Infrared Physics, Shanghai Institute of Technical Physics, Chinese Academy of Science, Shanghai 200083, People's Republic of China*

[2]*Key Laboratory of Polar Materials and Devices, Ministry of Education, East China Normal University, Shanghai 200062, People's Republic of China*

Yanfeng Wei,[3] Jianrong Yang,[3] and Li He[3]

[3]*Research Center for Advanced Materials and Devices, Shanghai Institute of Technical Physics, Chinese Academy of Sciences, 200083 Shanghai, China*

Magnetoconductance of a gated two-dimensional electron gas (2DEG) in the inversion layer on p-type HgCdTe crystal is investigated. At strong magnetic fields, characteristic features such as quantum Hall effect of a 2DEG with single subband occupation are observed. At weak magnetic fields, weak antilocalization effect in ballistic regime is observed. Phase coherence time and zero-field spin-splitting are extracted according to Golub's model. The temperature dependence of dephasing rate is consistent with Nyquist mechanism including both singlet and triplet channel interactions.

$Hg_{1-x}Cd_xTe$ (MCT) has narrow gap and strong spin-orbit interaction.[1] It not only dominates the infrared techniques for decades[1] but also harbors many interesting spin-related properties,[2] thus has potentials in spintronics.[3] In the triangular potential well formed in the inversion layers of narrow-band semiconductors, Rashba spin-orbit coupling dominates the spin-splitting due to the lack of structure inversion symmetry.[4,5,6] Rashba coupling has contributions from electric field as well as boundary conditions, it provides possibility to manipulate spin-orbit coupling electrically thus pave the way for spintronics.[4] Moreover, recent researches show that two-dimensional topological insulator[7,8,9] which has novel quantum spin Hall effect can emerge in $Hg_{1-x}Cd_xTe$ quantum wells when turning it into inverted-band regime by changing x.[9] Whether or not such topological insulator can be realized in the two-dimensional electron gas (2DEG) in the inversion layers of MCT is also an interesting problem.

For the investigation of Rashba spin-orbit coupling in MCT, most researchers employed the analysis of subband Landau levels through Shubnikov-de Haas (SdH) oscillations,[10,11] magneto-capacitance spectroscopy,[6,12,13] capacitance-voltage spectroscopy[10] or cyclotron resonance spectroscopy.[10,11] Some researchers analyzed beating-pattern in SdH oscillations.[2,14] Besides these methods using strong magnetic field, quite few researchers employed methods that work under weak magnetic field such as weak antilocalization effect (WAL).[15,16] WAL serves as a better tool for investigating spin-orbit interaction because (i) high-field spin splitting can change magnitude and/or sign therefore complicates the analysis of Rashba spin-splitting.[5] (ii) beating patterns in SdH oscillations may originate from inhomogeneities or a slightly occupied second subband.[17, 18]

WAL originates from interference of phase coherent electronic waves in the presence of spin-orbit interaction.[19] In heterostructures, spin-orbit interaction can be described by the following Hamiltonian:[20]

$$H(\boldsymbol{k}) = \hbar \boldsymbol{\sigma} \cdot \boldsymbol{\Omega}(\boldsymbol{k}), \tag{1}$$

where $\boldsymbol{k}$ is the electron wave vector, $\boldsymbol{\sigma}$ is the vector of Pauli matrices, $\boldsymbol{\Omega}(\boldsymbol{k})$ is an odd function of $\boldsymbol{k}$, $\hbar$ is the reduced Planck's constant. Most models concerning magnetoconductance caused by the destruction of WAL in magnetic field are valid for weak spin-orbit interaction and diffusion regime.[20] Weak spin-orbit interaction means $\Omega\tau \ll 1$,[20] where $\tau$ is transport scattering time. Diffusion regime exists for $B \ll B_{tr} = \hbar/2el^2$,[20] where $B$ is the magnetic field, $l$ is the mean free path. In high-mobility samples, both conditions can be violated. In addition, correction to conductivity from nonbackscattering interference in diffusion regime is not important, but it's very important in ballistic regime, thus calculations in different regimes are different.[20, 21] Golub developed a model that works at the whole range of classically weak magnetic fields and arbitrary strong spin-orbit interaction for 2DEG with only Rashba or Dresselhaus interaction.[20]

This work focuses on magnetoconductance of 2DEG with a singly occupied subband in the inversion layer of MCT. We observe weak antilocalization effect in ballistic regime at low magnetic fields. We extract phase coherence time and zero-field spin-splitting at various temperatures and gate voltages according to Golub's model. We find that the dominating dephasing mechanism is Nyquist mechanism including contributions from singlet as well as triplet channels.

The $5\times 5$ mm$^2$ large p-type Hg$_{1-x}$Cd$_x$Te film with x=0.218 used for the fabrication of our sample is grown on CdZnTe substrate by liquid-phase epitaxial (LPE) method. Anodic oxidation is used to form an inversion layer on the MCT film and a gate is made upon the inversion layer after deposition of rosin a few hundred μm thick as insulation layer [see inset of Fig. 1(b)]. Indium is used to facilitate Ohmic contacts. The magnetoresistance is measured in Van der Pauw configuration and the magnetic field is applied perpendicular to the film. A package of Keithley sourcemeters is used to measure magnetoresistance. All measurements are carried out in an Oxford Instruments $^4$He cryogenic system with temperature ranges from 1.3 to 9.0 K. At 1.4 K and zero gate-voltage, the carrier concentration is $3.0\times 10^{15}$ m$^{-2}$ and the mobility is 2.4 m$^2$V$^{-1}$s$^{-1}$.

In Fig. 1 (a) and (b), we can see excellent SdH oscillations and integer quantum Hall effect. Figure 1(b) shows $\rho_{xy}$ as a function of magnetic field, we can see quantum Hall plateaus which take on quantized values $2\pi\hbar/\upsilon e^2$ with $\upsilon$=1, 2, 3, 4. Figure 1(a) shows $\rho_{xx}$ as a function of magnetic field, we can see that resistance corresponding to $\upsilon$=2 vanishes. Carrier densities obtained from SdH oscillations and Hall measurements are very close. All of these phenomena are characteristic features of a high-quality 2DEG with a singly occupied subband which dominates the transport. The shift of SdH oscillations and change of $\rho_{xy}(B)$ curves are caused by the tuning of carrier density due to the gate.

The effective mass of ground electric subband is extracted by analyzing the temperature dependence of the amplitude $A(T)$ ($T$ is temperature) of the SdH oscillations.[22] The effective mass $m^*$ we get is $0.0185m_0$ ($m_0$ is the electron's mass in vacuum), this value is close to the effective mass of electrons at lowest subband in MCT inversion layers reported in other literatures.[23]

Clear weak localization effect with strong spin-orbit interaction has been observed in our sample. In Fig.2, we can see that a negative magnetoconductance is imposed on a positive magnetoconductance background. The positive magnetoconductance background is caused by weak localization effect. The negative magnetoconductance 'peak' before the emerging of positive magnetoconductance is a feature of strong spin-orbit interaction and is known as weak antilocalization effect.[19] With the increase of temperature, we can see that the negative magnetoconductance 'peak' fades away gradually.

For our sample, $B_{tr} \approx 7.4\,mT$ at zero gate-voltage. However, in Fig.2, we can see that the magnetic field at which weak antilocalization effect occurs spans at least from -5 to 5 mT, $B_{tr}$ is only slightly larger than the magnetic field at which the conductivity minimum happens. Therefore, the condition ($B \ll B_{tr}$) which justifies the application of weak antilocalization models in diffusion regime is not satisfied. Attempts of fitting experimental data with the widely-used ILP (Iordanskii, Lyanda-Geller and Pikus) model[24, 25] which is valid in diffusion regime give rise to poor fitting around the magnetic field at which the conductivity minimum happens. Weak antilocalization model in ballistic regime is appropriate here. After the subtraction of a parabola conductivity background which includes contributions from Drude conductivity and electron-electron interaction correction,[26] the quantum correction of the magnetoconductivity caused by WAL at zero gate-voltage is fit according to Golub's model:[20]

$$\Delta\sigma = \Delta\sigma_{xx}(B) - \Delta\sigma_{xx}(0) = \sigma_a(B, \tau/\tau_\varphi, \Omega\tau) + \sigma_b(B, \tau/\tau_\varphi, \Omega\tau) \qquad (1)$$

, where $\tau_\varphi$ is the phase coherence time. $\sigma_a$ and $\sigma_b$ are interpreted as the respective quantum corrections to conductivity from back-scattering and nonbackscattering interference.[20] The correction caused by nonbackscattering interference tends to reduce the magnitude of weak localization.[21] In diffusion regime, the contribution from nonbackscattering interference is not important, but in ballistic regime, the contribution from nonbackscattering interference is important and it should be taken into consideration.[20, 21] Forms and details about the calculation of $\sigma_a$ and $\sigma_b$ can be found in Ref. 27 and 20. In fitting procedure, the computed function $\Delta\sigma(B, \tau/\tau_\varphi, \Omega\tau)$ is stored as a matrix on a semilogarithmic mesh. Intermediate values between neighboring points $(B_i, \tau/\tau_{\varphi i}, \Omega\tau_i)$ and $(B_{i+1}, \tau/\tau_{\varphi i+1}, \Omega\tau_{i+1})$ are determined by spline interpolation.[27]

The fitting results are shown in Fig. 2. $\tau_\varphi$ ranges from $9.93\times 10^{-12}$ to $3.50\times 10^{-11}$ s. $\Omega\tau$ is basically temperature independent. $\tau_\varphi$ decreases as temperature increases, see Fig. 3. From the relation between $\tau_\varphi$ and $T$, we can get the information about the dephasing mechanism. Without magnetic impurities, phase relaxation comes from inelastic scattering caused by electron-phonon scattering or electron-electron scattering. Generally speaking, the temperature dependence of dephasing rate follows a power law, $\tau_\varphi^{-1} \sim T^p$. For electron-phonon scattering mechanism[28, 29], $p=2$ or 4. For electron-electron scattering mechanism (Nyquist dephasing mechanism) [30, 29, 31], $p=2$ at high temperature and $p=1$ at low temperature (because small-energy-tranfer process usually dominates at low temperature [29, 26]). For our data, the best fit is obtained for $p=1.12$ (see Fig.3, red line), thus the main dephasing mechanism is Nyquist mechanism.

In our sample, $k_B T \tau / \hbar \approx 0.07 \ll 1$ ($k_B$ is the Boltzmann's constant), this means that our sample is in the small-energy-transfer regime.[29, 26] In this regime, the simplest model about the Nyquist dephasing mechanism[30, 26] of Fermi liquid predicts a linear relation between $\tau_\varphi^{-1}$ and $T$:

$$\tau_\varphi^{-1} = \frac{1}{\alpha_N} \cdot T, \alpha_N = \frac{\frac{2\pi\hbar^2}{e^2}\sigma_0}{k_B \ln\left(\frac{\pi\hbar}{e^2}\sigma_0\right)} \qquad (2)$$

, where $e$ is the electronic charge, $\alpha_N$ is the dephasing time, $\sigma_0$ is zero-field conductivity.

According to parameters obtained from linear fitting, we have dephasing rate of zero gate-voltage $\alpha_N^{-1}$ ($2.74 \times 10^{10}$ s$^{-1}$), larger than theory value ($1.21 \times 10^{10}$ s$^{-1}$). The discrepancy indicates that Eq. (2) doesn't work very well here. In fact, this model only takes into consideration the singlet contribution,[26] a more complete model which considers both singlet and triplet contributions has been developed,[26, 32, 33, 29]

$$\tau_\varphi^{-1} = (1 + \frac{3(F_0^\sigma)^2}{(1+F_0^\sigma)(2+F_0^\sigma)})\frac{k_B T}{g\hbar}\ln(g(1+F_0^\sigma)) + \frac{\pi}{4}\left(1 + \frac{3(F_0^\sigma)^2}{(1+F_0^\sigma)^2}\right)\frac{(k_B T)^2}{\hbar E_F}\ln(\frac{E_F \tau}{\hbar}) \qquad (3)$$

, where $g = 2\pi\hbar\sigma_0 / e^2$ and $E_F$ is the Fermi energy. $F_0^\sigma$ is the interaction constant for the triplet channel, this parameter reflects the strength of spin-exchange interaction and can be obtained independently through means of measurement of the spin susceptibility of the system.[34] For instance, the spin-exchange interaction tends to align spins, even leads to ferromagnetic Stoner instability for $F_0^\sigma < 0$.[34] $F_0^\sigma$ has been investigated in various systems such as InGaAs quantum wells[29] or SiGe[35] quantum wells, and both negative and positive $F_0^\sigma$ have been observed.

The fitting according to Eq. (3) can be seen in Fig. 3 (green dash line), the best fit is reached when $F_0^\sigma = 0.616$, this positive $F_0^\sigma$ means that the spin-exchange interaction won't align spins in our sample.[34] We'll see that this value is quite close to the value obtained below in the analysis of data at varies gate voltages. Therefore the Nyquist mechanism for Fermi liquid when taking both singlet contribution and triplet contribution into consideration is a good description of dephasing mechanism in our sample.

After the application of a gate-voltage, in Fig. 4, we can see that the weak antilocalization effect can be tuned by the gate. It means that the spin-orbit coupling can be manipulated by a gate and thus a promising landscape concerning spintronic devices based on electrically manipulated spin-orbit coupling is possible.

$\Omega\tau$ and $\tau_\varphi$ at various gate-voltages are extracted according to Golub's model. From the dependence of $\tau_\varphi^{-1}$ upon $\sigma_0$, see Fig. 5, we can decide the better one for the description of dephasing between Eq. (2) and Eq. (3). Fitting according to Eq. (3) is the blue curve in Fig. 5, the resulting $F_0^\sigma = 0.617$ is quite close to the value (0.616) from the analysis of temperature dependence of $\tau_\varphi^{-1}$ (see Fig. 3). The red curve in Fig. 5 is a plot related to Eq. (2), we can see that Eq. (3) gives the correct dependence.

Thus the validity of Eq. (3) which takes into consideration both singlet contribution and triplet contribution in dephasing is further strengthened.

$\Omega\tau$ ranges from 0.416 to 0.542, the corresponding zero-field spin-splitting energy ($\Delta_0$) ranges from 2.4 to 2.8 meV, see Fig. 6(a). Compared with HgTe-based quantum wells which have inverted band, the obtained spin-splitting energy in our sample is not very large.[2] The Rashba parameter ($\alpha$) obtained from $\Omega\tau$ ranges from $9.93 \times 10^{-12}$ to $10.1 \times 10^{-12}$ eVm, see Fig. 6(b) The obtained Rashba parameter is close to values obtained in previous researches.[6, 15, 10, 14,36, 11] This $\alpha$ is quite large when compared with that of other systems.[17, 18, 37] This makes MCT a good candidate for the realization of spin-FET.[3] The large $\alpha$ is rooted in the relatively large ratio $\Delta_{so}/E_g$ (meaning of $\Delta_{so}$ and $E_g$) of MCT when compared with other materials.[14] Another factor influences the magnitude of $\alpha$ is the degree of the structure inversion asymmetry,[18] the relatively large structure asymmetry in the triangular potential well[38] in our sample can also help to develop a large Rashba parameter.[18]

The dependence of zero-field spin-splitting energy ($\Delta_0$) and $\alpha$ upon carrier density (*n*) can be seen in Fig. 6(a) and (b). The carrier density dependence of Rashba parameter is material-specific.[4, 17, 39] In most systems, the Rashba parameter either doesn't depend on carrier density or monotonically decreases as carrier density increases. For example, the carrier density independence of Rashba parameter in Al$_x$Ga$_{1-x}$N/GaN heterostructure can be attributed to the fact that the Rashba coupling is caused by the lack of crystal inversion symmetry in this system,[17] which is a crystal effect and cannot be controlled by a gate. In contrast, the Rashba coupling in inversion layers of our sample as well as other narrow-band semiconductors comes from the lack of structure symmetry in the triangle potential well,[4, 5, 6] it is correlated with carrier density and electric field in the well. The possibility of nonmonotonic dependence of $\alpha$ upon *n* in MCT has been investigated theoretically, a mechanism concerning the decrease of the interband coupling due to the increase of the kinetic energy of the subband has been proposed.[40] Further experimental investigation is necessary in order to clarify the *n* dependence of $\alpha$.

In summary, weak antilocalization effect in the inversion layer of HgCdTe film is observed. After fitting the data at varies temperatures and gate voltages with Golub's model, phase coherence time and zero-field spin-splitting energy are extracted as a function of temperature and gate voltage. We find that Nyquist dephasing mechanism is the dominating dephasing mechanism.

**Acknowledgements**

Authors of this work thank Longyuan Zhu, Guangyin Qiu for their help in the processing of samples. This work was supported in part by the Special Funds for Major State Basic Research under Project No. 2007CB924901, National Natural Science Foundation of China under Grant Nos. 60976093 and 60906045, Science and Technology Commission of Shanghai under Grant No. 09JC1415700.

**Figure captions**

FIG. 1. (a) Magnetoresistance of our sample at different gate voltages. (b) Hall resistance at various gate voltages. Inset shows the configuration of our sample, region A (white) is p-type MCT film, region B (blue) is the inversion layer, region C (black) is electrode, region D (yellow) is the insulation layer, region E (brown) is the gate.

FIG. 2. Weak antilocalization effect of our sample at different temperatures with no gate-voltage applied. Blue circles are the experimental data and red curves are fitting according to Golub's model. The curves and circles have been shifted vertically for clarity.

FIG. 3. $\tau_\varphi^{-1}$ (blue circles) corresponding to zero gate-voltage as a function of temperature. Red curve is fitting result according to $\tau_\varphi^{-1} \sim T^p$, green dash line is fitting according to Eq. (3).

FIG. 4. Weak antilocalization effect at different gate voltages at 1.4 K. Blue circles are the experimental data and red curves are fitting according to Golub's model. The curves and circles have been shifted vertically for clarity.

FIG. 5. $\tau_\varphi^{-1}$ as a function of $\sigma_0$. Green blades are the experimental data and blue curve is fitting according to Eq. (3); red curve is a plot related to Eq. (2).

FIG. 6. (a) Zero-field spin-splitting energy $\Delta_0$ as a function of carrier density, blue triangles are experimental data. (b) Rashba parameter $\alpha$ as a function of carrier density.

Figure 1

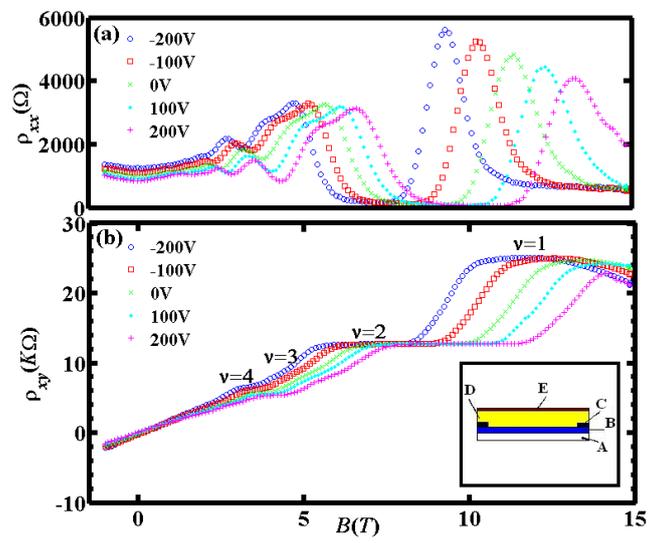

Figure 2

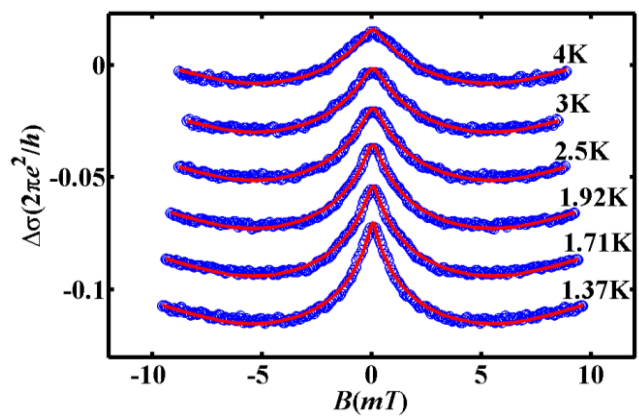

Figure 3

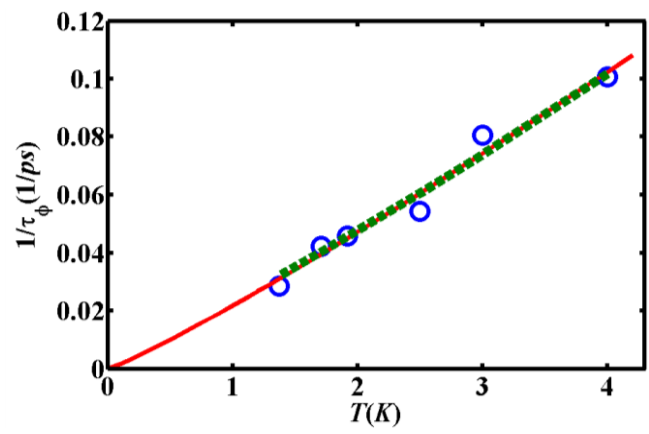

Figure 4

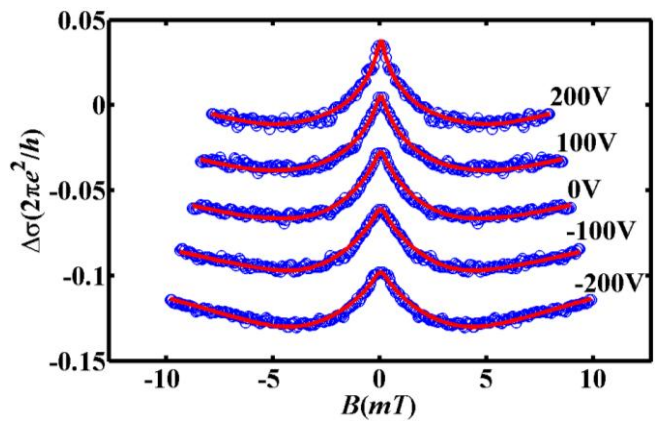

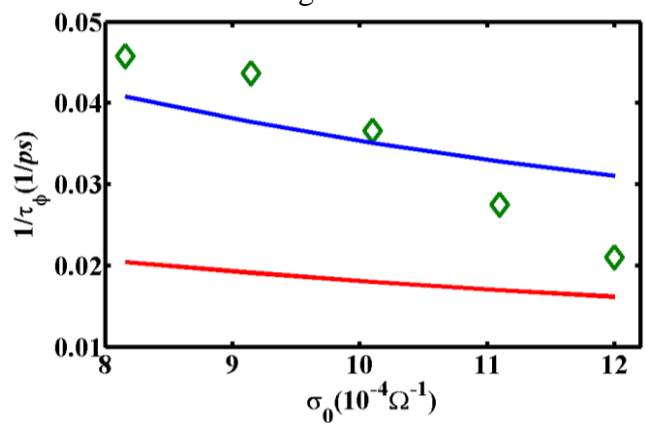

Figure 5

Figure 6

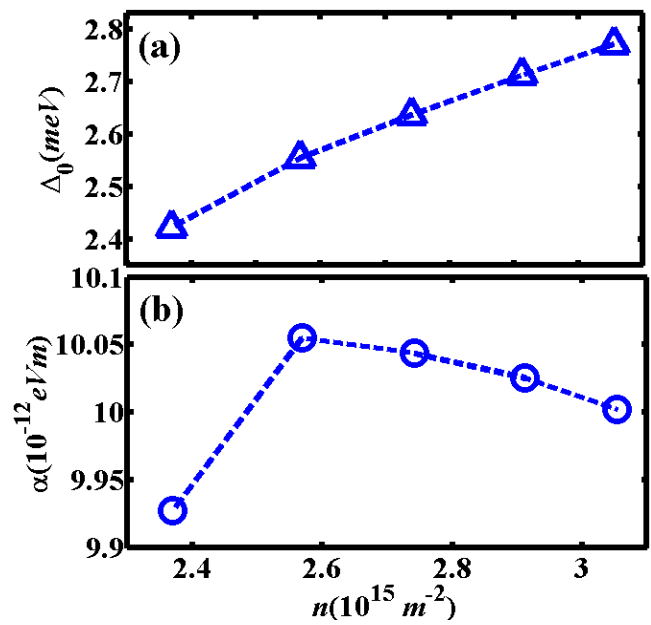